\documentclass[aps,prl,reprint,superscriptaddress,floatfix]{revtex4-2}

\usepackage{graphicx}
\usepackage[separate-uncertainty=false]{siunitx}
\usepackage{caption}
\usepackage{lineno}
\usepackage{framed}
\usepackage{siunitx}
\usepackage{ragged2e}
\usepackage{multirow}
\usepackage{amsmath}
\usepackage{braket}
\begin{document}

\title{Continuous variable distributed quantum sensing in integrated photonics}
\author{Bethany~Puzio}
\affiliation{Quantum Engineering Technology Labs, H. H. Wills Physics Laboratory and School of Electrical, Electronic, and Mechanical Engineering, University of Bristol, BS8 1FD, UK}
\affiliation{Quantum Engineering Centre for Doctoral Training, H. H. Wills Physics Laboratory and School of Electrical, Electronic, and Mechanical Engineering, University of Bristol, BS8 1FD, UK}

\author{Oliver~M.~Green}
\affiliation{Quantum Engineering Technology Labs, H. H. Wills Physics Laboratory and School of Electrical, Electronic, and Mechanical Engineering, University of Bristol, BS8 1FD, UK}
\affiliation{Quantum Engineering Centre for Doctoral Training, H. H. Wills Physics Laboratory and School of Electrical, Electronic, and Mechanical Engineering, University of Bristol, BS8 1FD, UK}

\author{Joel~F.~Tasker}
\affiliation{Quantum Engineering Technology Labs, H. H. Wills Physics Laboratory and School of Electrical, Electronic, and Mechanical Engineering, University of Bristol, BS8 1FD, UK}

\author{Jonathan~Frazer}
\affiliation{Quantum Engineering Technology Labs, H. H. Wills Physics Laboratory and School of Electrical, Electronic, and Mechanical Engineering, University of Bristol, BS8 1FD, UK}

\author{Tamzin~Ellis}
\affiliation{Quantum Engineering Technology Labs, H. H. Wills Physics Laboratory and School of Electrical, Electronic, and Mechanical Engineering, University of Bristol, BS8 1FD, UK}
\affiliation{Quantum Engineering Centre for Doctoral Training, H. H. Wills Physics Laboratory and School of Electrical, Electronic, and Mechanical Engineering, University of Bristol, BS8 1FD, UK}

\author{Benjamin~D.~J.~Sayers}
\affiliation{Quantum Engineering Technology Labs, H. H. Wills Physics Laboratory and School of Electrical, Electronic, and Mechanical Engineering, University of Bristol, BS8 1FD, UK}

\author{Rachel~N.~Clark}
\affiliation{Quantum Engineering Technology Labs, H. H. Wills Physics Laboratory and School of Electrical, Electronic, and Mechanical Engineering, University of Bristol, BS8 1FD, UK}

\author{Alex~S.~Clark}
\email[]{alex.clark@bristol.ac.uk}
\affiliation{Quantum Engineering Technology Labs, H. H. Wills Physics Laboratory and School of Electrical, Electronic, and Mechanical Engineering, University of Bristol, BS8 1FD, UK}

\author{Giacomo~Ferranti}
\affiliation{Quantum Engineering Technology Labs, H. H. Wills Physics Laboratory and School of Electrical, Electronic, and Mechanical Engineering, University of Bristol, BS8 1FD, UK}

\author{Jonathan~C.~F.~Matthews}
\affiliation{Quantum Engineering Technology Labs, H. H. Wills Physics Laboratory and School of Electrical, Electronic, and Mechanical Engineering, University of Bristol, BS8 1FD, UK}

\date{\today}% It is always \today, today,
% abstract below is over 150 word limit - have rewritten 
% \begin{abstract}
% Distributed quantum sensing is an emerging application of quantum networking, where entangled probe states are employed to sense combinations of delocalized parameters with an enhanced precision relative to using separable states. Squeezed states of light have proven to be prime resource for experimental demonstrations of entanglement-enhanced sensing, because they can be generated and entangled deterministically using linear optics. Existing distributed quantum sensing experiments have been fundamentally limited in scalability due to their bulk-optic architectures. Meanwhile integrated photonics provides a sufficiently scalable and compact alternative platform for quantum sensors. Here we demonstrate the entanglement-enhanced sensing of linear functions of four phase shifts in an integrated photonic circuit. We find an entanglement-enhanced precision up to 0.224 dB below the shot nose limit, while separable squeezed states only demonstrate a precision up to 0.063 dB below the shot noise. The entanglement is generated on-chip using an integrated interferometer network, with entanglement verification and phase sensing also performed on-chip with an array of four integrated homodyne detectors.
% \end{abstract}
%TC:ignore
\begin{abstract} \bfseries \boldmath
% Start with one or two sentences of background
    Distributed quantum sensing is an emerging application of quantum networking, where entangled probe states are employed to sense combinations of delocalized parameters with enhanced precision relative to using separable states. Squeezed states of light are a prime resource for experimental demonstrations of entanglement-enhanced sensing, because they can be generated and entangled deterministically. Existing distributed quantum sensing experiments have been fundamentally limited in scalability due to their bulk-optic architectures. Meanwhile, integrated photonics provides a scalable and compact platform for quantum sensors. Here we demonstrate entanglement-enhanced sensing of linear functions of four phase shifts in an integrated photonic circuit. We find an entanglement-enhanced precision of 0.199(16)~dB below the shot noise limit compared to 0.041(18)~dB for separable states. A four-mode entangled state is generated on-chip with entanglement verification and phase sensing also performed on-chip with an array of four integrated homodyne detectors.
\end{abstract}
\maketitle

% The first paragraph of any Science paper does NOT have a heading
% Nor is it indented
\noindent
%TC:endignore
\section{Introduction}
\label{sec:Introduction}
%Current word count: ~440

Quantum networking and sensing are two rapidly advancing use cases of technologies able to harness fundamental quantum mechanical properties to gain an advantage that is unattainable in classical systems. In quantum networking, distributing entanglement between distant nodes permits secure communications between parties \cite{Kimble2008} and scaling quantum computation modules \cite{Main2025}. In quantum sensing, physical parameters may be measured with a precision beyond the limits imposed by classical physics strictly when using quantum probe states \cite{Giovannetti2011}, with use cases for optical sensing including sub-shot noise imaging of biological samples \cite{Nasr2009,Taylor2013,Taylor2014}, high precision atomic clocks \cite{Letargat2013,Bloom2014}, and gravitational wave detection \cite{Caves1981,Tse2019}. Distributed quantum sensing (DQS) is an emerging application that combines these two fields. Sensing schemes require probe states to initially interact with a system of interest, and subsequent measurement outcomes of the probe allow construction of an estimator from which we infer parameters related to  the system of interest \cite{Giovannetti2004}. In a DQS scheme, multiple entangled probes perform a sensing protocol over multiple sites simultaneously to sense a global parameter. Theoretical investigations highlight key applications for DQS, including atomic clock synchronisation \cite{Komar2014}, molecular tracking \cite{Qi2018} and privacy networks in which parties have access to global measurement but not individual parameters \cite{Shettell2022}. DQS schemes achieve higher precision in global parameter estimation when compared to individual quantum probe states that do not share entanglement and simultaneously sense local parameters to infer a global property. Importantly, DQS schemes are distinct from multi-parameter estimation protocols, which estimate multiple local parameters rather than a single global parameter. In particular, theoretical analyses for optical schemes have identified the presence of entanglement-enhanced measurement precision with estimators formed from linear combinations of displacement and phase measurements \cite{Ge2018, Zhuang2018, Gatto2019, Oh2020, Oh2022, Malitesta2023}. This subsequently prompted experimental demonstrations in both the discrete variable (DV) \cite{Liu2020, Zhao2021, Kim2024, Ho2026} and the continuous variable (CV) regime \cite{Guo2019, Xia2020, Brady2023}, where single photons and squeezed states of light are the primary resources, respectively. Squeezed light is characterised by the presence of below-vacuum noise at specific phases, offering a sub-shot noise measurement precision which enables optical quantum sensing. In addition, CV entanglement can be generated deterministically with a multi-mode beamsplitter network (BSN), in contrast to linear optical DV entanglement generation which is inherently probabilistic with linear optics. Single photon measurement often relies on cryogenic detectors, whereas squeezed state measurement uses optical homodyne detection based on room temperature photodiodes. Together these properties make the CV regime particularly appealing for realising DQS schemes. 

\begin{figure*}%[!htb]
    \centering
    \includegraphics[width = \textwidth]{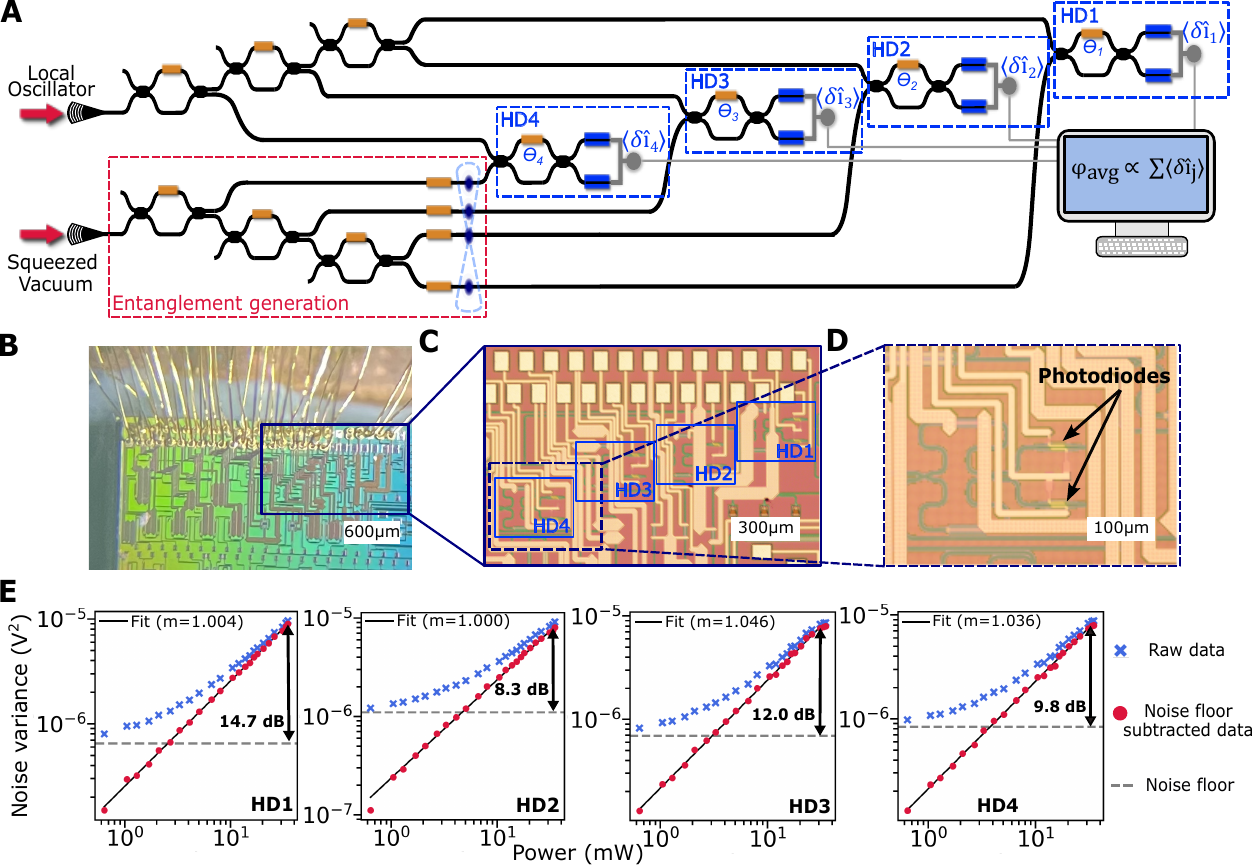}
    \caption{\textbf{The photonic chip} (A) Layout of the silicon-on-insulator (SOI) photonic integrated circuit (PIC). The single mode squeezed vacuum and local oscillator (LO) are coupled onto the PIC via grating couplers. Entanglement is generated across four spatial modes by mixing the squeezed light with vacuum across three cascaded Mach-Zehnder interferometers (MZIs). The LO is also split across four spatial modes via MZIs, and then routed to homodyne detectors HD1-4, to detect each mode of the entangled state. Thermo-optic phase modulators (TOPMs) $\theta_{1-4}$ are used to balance HD1-4 respectively, while also acting as the phase parameters to be sensed. (B) Optical microscope image of the PIC and wirebonds, with higher resolution images of (C) the array of four homodyne detectors and (D) HD4 with the germanium photodiodes labelled. (E) Simultaneous shot noise clearance measurements of HD1-4 with the maximum clearance and linear fit gradients labelled.
    }
    \label{fig:1}
\end{figure*}

DQS experiments have been demonstrated with bulk optics which occupy a large footprint and are sensitive to environmental disturbances, restricting the ability to deploy and scale these quantum sensors. Recent advances in nanofabrication technology mean that photonic integrated circuits (PICs) offer a full component toolkit for a scalable and phase-stable CV DQS scheme. Integrated optical homodyne detectors have also been demonstrated on chip making use of on-chip beam splitters and high responsivity waveguide-integrated germanium photodiodes \cite{Raffaelli2018,Tasker2020,bruynsteen_integrated_2021,Yin2021,Jia2023,Tasker2024}. CV PICs have been developed for use in a variety of quantum technology applications \cite{Zhang2019, Jia2025, Jia2026,Xanadu2025,Clark2026}, with the exception to date of DQS schemes.

%This allows for multi-mode BSNs to be condensed to the millimetre-scale \cite{Carolan2015,Wang2018,Bao2023}. 
%homodyne detectors are formed from these on chip beam splitters and wavegeuige coupl with on-chip beamsplitters based on multi-mode interferometers (MMI) \cite{Raffaelli2018,Jia2023} or MZIs for tunable balancing \cite{Tasker2020, Yin2021,Tasker2024,Bian2025}, and outputs coupled into high efficiency germanium photodiodes..

In this work, we harness the advantages of CV  quantum integrated photonics to perform a DQS experiment in which we sense linear functions of phase parameters with entanglement-enhanced precision. The backbone of this experiment is a silicon-on-insulator (SOI) PIC featuring an array of four integrated homodyne detectors, as well as an integrated BSN which enables the deterministic generation of four-mode entanglement from an off-chip source of squeezed vacuum. The BSN is fully reconfigurable, allowing for arbitrary routing of the incoming squeezed vacuum to produce two-, three-, or four-mode entanglement. The four homodyne detectors provide a means of interrogating the multi-mode state for characterisation and the sensing scheme. %In addition to balancing, t
%The TOPM in the MZI in each homodyne detector is used to encode the phase to be sensed. 
Together, using these key components, we report the first example of a DQS protocol implemented on a foundry-compatible chip. In the following sections, we first present the SOI PIC and the individual components used for the DQS protocol, before detailing the implementation of the phase sensing protocol. We then outline the noise-locking procedure to continually interrogate the squeezed quadrature of the probes, and the scheme for verifying four-mode entanglement on chip. Finally we detail the results of the DQS protocol, demonstrating an entanglement-enhanced precision on the global phase parameter across the four spatial modes.

\section{Results}
\subsection{The photonic integrated circuit and sensing scheme}
The SOI PIC utilised for this experiment was fabricated by IMEC Foundry Services on a multi-project wafer using their iSiPP50G process, with a chip schematic shown in Fig.~\ref{fig:1}(A). Thermo-optic phase modulators (TOPMs) are employed as phase shifters on MZIs to construct beam splitters of arbitrary reflectivity and on individual waveguides to impart separate phase shifts. An off-chip source of squeezed vacuum (SV) couples onto the PIC via a grating coupler. The SV enters a BSN formed of three cascaded MZIs, and subsequently a multipartite entangled state is generated by distributing the input SV across four spatial modes. Each spatial mode is then routed to an integrated homodyne detector where it is interfered with a coherent local oscillator (LO) for state measurement. %and phase parameter sensing.
The TOPMs at each homodyne detector ($\theta_{1-4}$, $\text{HD}_{1-4}$ in Fig.~\ref{fig:1}) are initially to `balance' the homodyne detector, ensuring an optimal common-mode rejection ratio. The phase angle of the measured quadratures is set by the relative phase between the squeezed state and the LO, which is controlled by TOPMs on the paths of the multipartite squeezed state after the BSN. 
The output of each homodyne detector is wire-bonded to a custom-made printed circuit board (PCB) where each subtracted photocurrent is amplified and converted to a voltage through a discretely packaged transimpedance amplifier (TIA) (see Supplementary Materials for further details). Figure~\ref{fig:1}(B) shows an image of the PIC and wirebonds. Fig.~\ref{fig:1}(C) shows a microscope image with the location of each of the homodyne detectors labelled, and Fig.~\ref{fig:1}(D) shows a further zoomed in microscope image of the fourth homodyne detector (HD4) with the germanium photodiodes labelled. 

The sensing scheme is implemented as follows. When the TOPMs at each homodyne detector are set in the balanced configuration, any small phase shift from this point will result in a direct current (DC) offset of the output. The overall phase set on the MZI at each homodyne detector ($\theta_{1-4}$, $\text{HD}_{1-4}$ in Fig.~\ref{fig:1}) can therefore be expressed as $\theta_{j}=\phi_{j}+\varphi_{j}$ for $j\in\{1,2,3,4\}$ where $\phi_{j}$ is the phase set by the TOPM to balance the detector and $\varphi_{j}$ is the small additional phase shift to be measured. Measuring the mean subtracted photocurrent $\braket{\delta \hat{i}_{j}}$ at each detector reveals $\varphi_{j}$ as, $\braket{\delta \hat{i}_{j}} = |\alpha_{\textrm{LO}}|^{2}\varphi_{j}$, where $|\alpha_{\textrm{LO}}|$ defines the LO coherent amplitude. When measuring the average of $\varphi_{j}$ over $M$ spatial modes, the phase sensing estimator can be constructed as

\begin{equation}
    \begin{split}
        \braket{\delta\hat{i}_{\mathrm{avg}}}
        &= \frac{1}{M}\sum_{j=1}^{M}\braket{\delta\hat{i}_{j}} \\
        &= |\alpha_{\textrm{LO}}|^{2}\varphi_{\mathrm{avg}} . \\
    \end{split}
    \label{eq:avg_subtracted_photocurrent_main}
\end{equation}

Using standard error propagation, the precision with which $\varphi_{\mathrm{avg}}$ is estimated by an $M$-mode entangled state generated by splitting a single mode squeezed vacuum state equally over an $M$-mode BSN is 

\begin{equation}
\begin{aligned}
\sigma_{e}
&= \frac{1}{|\alpha_{\textrm{LO}}|\sqrt{M}} \Bigg[
\left(\frac{1}{M}\sum_{j=1}^{M}\varphi_{j}^{2}\right) \\
&\quad + \frac{1}{2}\left(
\frac{\eta}{(\sqrt{N+1}+\sqrt{N})^{2}}
+ (1-\eta)
\right)
\Bigg]^{1/2}.
\end{aligned}
\label{eq: standard_error_propagation_ent_main}
\end{equation}

Here, $N$ is the mean photon number in the system such that $N=n_{sq}M$, where $n_{sq}$ is the mean photon number per mode. The achievable precision using $M$ separable squeezed probe states, $\sigma_{s}$, can be written in the same form as $\sigma_{e}$, however in this case $N=n_{sq}$. Therefore the entangled sensing protocol exhibits a $1/\sqrt{M}$ precision advantage over the 
\begin{figure*}%[htb!]
    \centering
    \includegraphics[width = \textwidth]{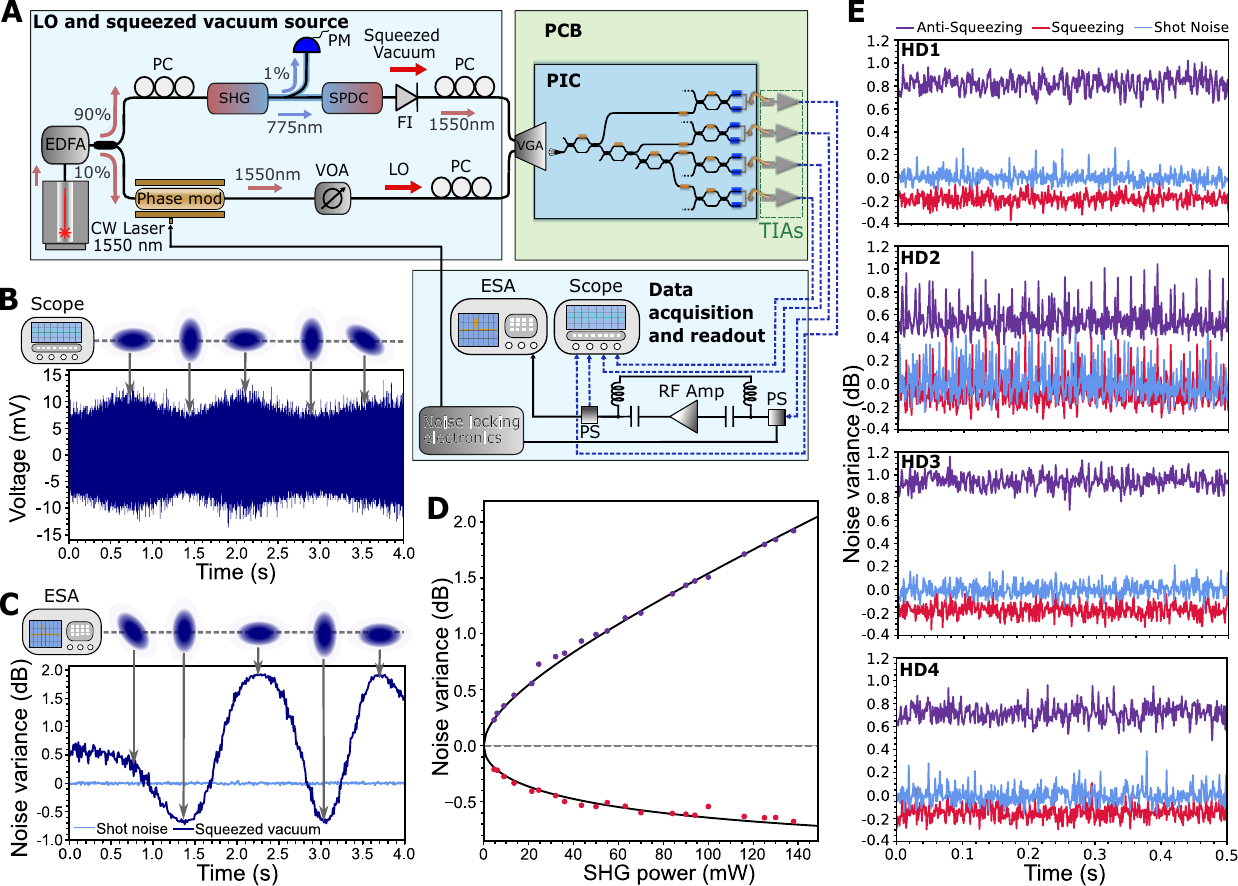}
    \caption{\textbf{Multi-mode squeezing measurements with integrated homodyne detectors} (A) Experiment schematic. LO and squeezed vacuum couple onto PIC via VGA. Each homodyne detector  output is wirebonded to a TIA mounted on a PCB. All TIA outputs are measured on the oscilloscope (scope). HD2 output is split into three paths using power splitters (PS) and an RF amplifier (RF amp) to derive the locking signal whilst measuring output simultaneously on the ESA and the oscilloscope. See Supplementary Materials for noise locking electronics details. (B) Squeezed vacuum noise measured on an oscilloscope with the corresponding phase space representation for illustration of the behaviour at specific points in time. Also plotted on the same timescale, (C) shows squeezed vacuum measurements on HD2 with the ESA in zero span mode. (D) Measured squeezing and antisqueezing levels as a function of SHG pump power. (E) Squeezed vacuum split across all four spatial modes with the locking protocol activated to lock to either the squeezed or anti-squeezed quadrature at all four detectors (HD1-4) simultaneously. All measurements of squeezing are normalised to shot noise level.
    }
    \label{fig:2}
\end{figure*}
separable sensing protocol when the average photon number per mode is fixed (see Supplementary Materials for further details). 
%For this experimental scheme, which uses an off-chip source of SV, this requires a noise-locking protocol.
% requires measurement along the squeezed access and so we impliment a protocol to lock the phase of the LO fopr homodyne detection to the SV.
\subsection{Squeezed vacuum source and noise locking protocol}
%~520 words
The source of SV is off-chip and fibre based, constructed from two commercially available periodically poled lithium niobate (PPLN) waveguides (NTT Innovative Devices) in an optical parametric amplifier (OPA) configuration \cite{Tasker2020,Tanzilli2016} (Fig.~\ref{fig:2}(A)).
An amplified continuous wave (CW) laser is passed into the first PPLN crystal for second harmonic generation (SHG). The output then pumps a spontaneous parametric down conversion (SPDC) process in the second PPLN crystal to generate squeezed vacuum. A portion of the original CW laser is retained as the LO. Further details of the SV source characterisation can be found in the Supplementary Materials.
% Move this to supplement
%A tunable continuous wave (CW) laser is first amplified with an erbium-doped fibre amplifier before being split into two paths by a 90:10 beam splitter. The $10\%$ port is reserved as the LO while the remainder is used to generate the squeezed light by spontaneous parametric down conversion (SPDC). This is facilitated by using the first PPLN crystal to frequency-double the signal to $\SI{775}{nm}$, with this output driving the SPDC in the second PPLN crystal. This configuration allows the LO and SV to be derived from the same source such that they have a fixed phase relationship and can be interfered for homodyne detection once coupled onto the PIC. 
Figure~\ref{fig:2}(B) shows a measurement of the HD TIA output on an oscilloscope (Keysight Infiniium EXR608A) when all squeezed vacuum is routed to one of the integrated detectors while Fig.~\ref{fig:2}(C) shows the same measurement configuration on an electronic spectrum analyser (ESA) in zero span mode (Keysight N9020B, 75 MHz central frequency, 8 MHz RBW). Using the ESA, and with all squeezing routed to a single detecor, we can extract the levels of on-chip squeezing as a function of SHG pump power, as shown in Fig.~\ref{fig:2}(D). At the maximum obtainable SHG pump power of \SI{350}{\milli\watt} we obtained \SI{0.84}{\decibel} of measured squeezing on-chip. From the fit of Fig.~\ref{fig:2}(D) we extract a total end-to-end system efficiency of $\eta=0.203(5)$. This includes a significant \SI{2.5}{\decibel} loss contribution from the grating coupler in addition to loss incurred in the fibres, PIC and detector inefficiency (see Supplementary Material for further details). With the squeezed vacuum and LO travelling through separate fibres prior to the PIC, thermal phase fluctuations between these paths lead to relative phase rotations preventing sustained measurement along the squeezed axis as shown in Fig.~\ref{fig:2}(B) and (C), with the corresponding rotations in phase space labelled. To implement our sensing scheme, and obtain the quadrature statistics needed to construct a multi-mode covariance matrix for entanglement verification, prolonged measurements of the squeezed and anti-squeezed quadratures are required. To overcome the innate thermal phase drift between the SV and LO, we choose to implement the noise locking protocol detailed by McKenzie \textit{et al.} \cite{McKenzie2005} (see Supplementary Materials for further details).
%The output of HD2 is directed to a chain of power splitters (PS) and RF amplifiers to derive the noise locking signal while also monitoring the locking on the ESA (Fig. ~\ref{fig:2}D, top) and collecting data for entanglement verification and sensing on the oscilloscope. Fig. ~\ref{fig:2}D (centre) shows the resulting ESA trace when all the squeezing is directed to HD2 and the noise locking protocol is activated to lock to the squeezed quadrature. The spike at around 3 seconds arises from the lock momentarily destabilising. Fig. ~\ref{fig:2}D (bottom) shows the lock switching from the squeezed quadrature to the anti-squeezed at around 1.5 seconds which is achieved by inverting the sign of the error signal entering the PID controller. 
Once the SV and LO have been coupled onto the PIC, the paths they take are inherently phase stable. Therefore we can use the locking signal derived from HD2 to globally lock to either the squeezed or anti-squeezed quadrature on all four modes simultaneously. Where each spatial mode has a slightly different path length prior to detection, we use TOPMs on the paths of the multipartite squeezed state after the BSN (Fig. \ref{fig:1}(A)) to adjust their relative phases. Fig.~\ref{fig:2}(E) shows the resulting measurements of locking to either the squeezed or anti-squeezed quadrature on all four modes simultaneously, normalised to the shot noise level. 
%Move to supplement
%This data was collected on the oscilloscope such that the DC offset can be retained for the phase sensing application on each mode. By collecting 100 million data points over 0.5 seconds, each variance value corresponds to the variance of a \SI{1}{\milli\second} temporal bin. 
Here we measure \SI{0.184}{\decibel}, \SI{0.080}{\decibel}, \SI{0.177}{\decibel}, and \SI{0.145}{\decibel} of squeezing on HD1, HD2, HD3 and HD4 respectively, where the lower squeezing level on HD2 is attributed to the presence of noise introduced by the additional RF electronics required to also derive the locking signal from this detector, visible in the noise variance measurements made on HD2.  

\subsection{Entanglement verification and phase estimation}
% ~900 words
The majority of existing CV entanglement demonstrations verify entanglement at a specific sideband frequency \cite{Guo2019,Larsen2019,Jia2026,Jia2025,Masada2015}, enforcing a well-defined frequency mode upon the quadratures which is free from low frequency technical noise. However, we must use time-domain data to preserve the DC offset and therefore phase information, necessitating a careful choice of temporal mode. Quadrature measurements are then extracted by integrating the output photocurrent over this chosen temporal mode \cite{kumar2012}. 
%While we are generally free to choose this temporal mode, since the phase sensing protocol presented in this work relies on extracting a DC signal, the selected temporal mode in the frequency domain must preserve the required phase information.
%ensure that temporal mode in the fourier domain mainitinas a DC componenet and that it must also map onto a mode where there is lots of squeezing (ie not much techniocal nosie)  However we must use time-domain data to preserve the DC offset and therefore phase information, necessitating a careful choice of temporal mode imposed on the quadratures. 

\begin{figure*}%[h!]
    \centering
    \includegraphics[width = \linewidth]{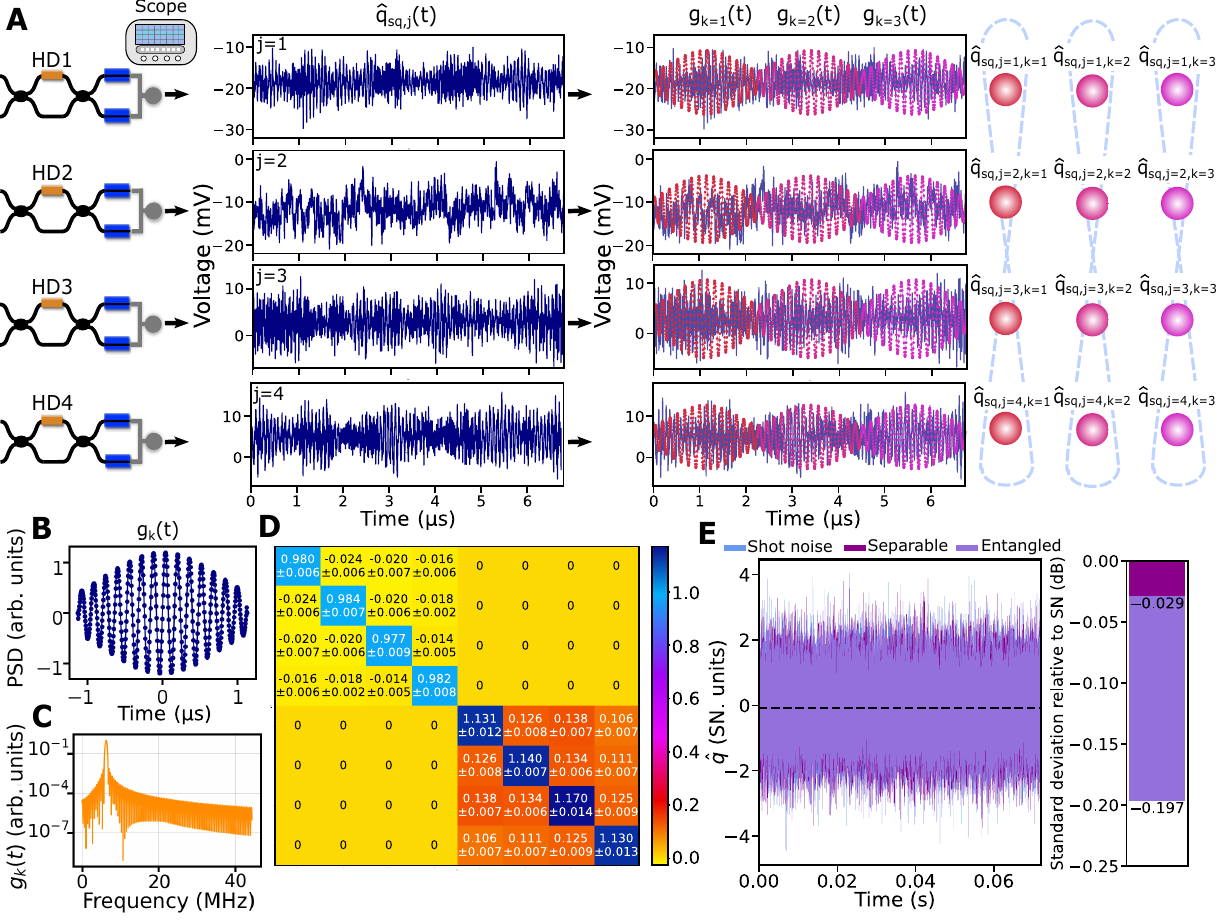}
    \caption{\textbf{Temporal mode definition and multi-mode state tomography} (A) Schematic illustrating the process of converting raw oscilloscope squeezing data, $\hat{q}^{\textrm{det}}_{sq,j}(t)$, into well-defined quadratures, $\hat{q}_{sq,j,k}$, where $j$ and $k$ define the spatial and temporal mode respectively. (B) Our choice of temporal mode function in time, $g_{k}(t)$, a sine wave modulated Gaussian with total width $\tau=2.25$ \si{\micro\second}, full width half maximum $s=1.8$ \si{\micro\second} and modulation frequency, $f=6.296$ MHz. (C) Temporal mode profile in frequency as a Gaussian centered on $f$. Crucially, the DC offset at 0 Hz is not completely suppressed. (D) Resulting four-mode covariance matrix and errors from 8 repeated tomography measurements of the quadratures acquired using the process shown in (A). (E) An example trace of the average shot noise, separable and entangled quadratures across the four spatial modes, with the quadrature mean marked by the dashed black line. The bar chart shows the standard deviation of the separable and entangled traces relative to the shot noise.}
    \label{fig:3}
\end{figure*}

A single quadrature value corresponding to the $k^{\text{th}}$ well-defined temporal mode, $\hat{q}_{k}$, is obtained by integrating the measured quadratures as a function of time, $\hat{q}(t)$, over the weight function defining the temporal mode $f_{k}(t)$. However, we must not neglect that the homodyne detector maps the optical signal into the electronic domain in accordance with its response function. This means that instead of directly measuring $\hat{q}(t)$, the homodyne detector outputs quadratures as $\hat{q}^{\textrm{det}}(t)$. We must therefore choose a computational weight function, $g_{k}(t)$, such that we obtain $\hat{q}_{k}$, as,
\begin{equation}
    \hat{q}_{k}=\int f_{k}(t)\hat{q}(t)\textnormal{d}t=\int g_{k}(t)\hat{q}^{\textrm{det}}(t) \textnormal{d}t .
    \label{eq: quadrature definition}
\end{equation}
An optimal choice of $g_{k}(t)$ is chosen via a routine which optimises conditionally based on the level of squeezing, the resulting logarithmic negativity (the entanglement witness for this scheme), and the physicality of the covariance matrix (see Supplementary Materials for further details). The process of obtaining the quadrature associated with the temporal mode $g_{k}(t)$ measured on the $j^{\text{th}}$ detector is visualised by the schematic in Fig. \ref{fig:3}(A). The output of each of the four detectors is simultaneously measured on the oscilloscope over a period of time to produce the data trace $\hat{q}^{\textrm{det}}_{sq, j}(t)$. The optimised temporal mode profile for the first three temporal modes is shown overlaid on $\hat{q}^{\textrm{det}}_{sq, j}(t)$ in the right of Fig. \ref{fig:3}(A). Each data trace is then integrated with this weight function over the region covered by each successive temporal mode to produce the quadratures $\hat{q}_{sq,j,k}$. The presence of entanglement between quadratures on spatial modes $j\in\{1,2,3,4\}$ that are aligned in time at temporal moment $k$ is highlighted by the dashed blue lines. 
Our choice of $g_{k}(t)$ is shown in Fig. \ref{fig:3} in both the (B) time and (C) frequency domain. We note the presence of a DC offset in the voltage versus time oscilloscope data which corresponds to the small phase offset that we will sense, and that the choice of $g_{k}(t)$ therefore preserves this signal.

The process shown in Fig. \ref{fig:3}(A) was repeated for both the anti-squeezing $\hat{q}^{\textrm{det}}_{asq,j}(t)$ and the shot noise $\hat{q}^{\textrm{det}}_{sn,j}(t)$, to measure the required statistics for the state's four-mode covariance matrix which is given in Fig. \ref{fig:3}(D). Entanglement verification was performed using the logarithmic negativity as a witness (see Supplementary Materials for further
details). On a system of $M$ modes, the bipartite entanglement for the logarithmic negativity must be partitioned as 1 vs $m-1$, where $m$ is some subset of the total number of modes. Therefore we cannot directly verify entanglement on the four moded state. Instead, we infer multimode entanglement by calculating the logarithmic negativity for all possible 1 vs $M-1$ mode bipartitions \cite{Serafini2017}. Specifically, we examine all 1 vs. 1, 1 vs. 2 and 1 vs. 3 mode combinations and obtain an average logarithmic negativity of $0.010(5)$, $0.017(5)$ and $0.022(5)$ for each case respectively. To confirm these positive logarithmic negativity values arise from entanglement and not classical correlations, we reconstruct the covariance matrix, shifting the data on each of the four modes by one temporal mode with respect to each other. We then find that all measured logarithmic negativity values are zero as the cross-spatial mode entanglement is destroyed. With these measurements confirming the existence of quadrature entanglement across all examined mode combinations, we conclude that four mode entanglement must be present. 
%Then, using the logarithmic negativity as a witness, entanglement verification was performed between all possible 1 vs. $M-1$ mode bipartitions . For all possible 1 vs. 1, 1 vs. 2 and 1 vs. 3 mode combinations, an average logarithmic negativity of $0.01\pm 0.005$, $0.017\pm 0.005$ and $0.022\pm 0.005$ respectively was measured, confirming the existence of quadrature entanglement across all possible mode combinations.

With cross-spatial mode entanglement verified, the DC offset contained within the quadratures was used to implement our DQS protocol. The individual phase $\varphi'_{j}$ on each mode $j$ can be sensed using 
\begin{equation}
    \braket{\hat{q}_{sq,j,k}}_{k}=A_{j}\varphi'_{j}+v_{j,0},
    \label{eq:AmplifiedPhaseEstimator}
\end{equation}
%$\braket{\hat{q}_{sq,j,k}}_{k}=A_{j}\phi'_{j}+v_{j,0}$ 
where $A_{j}$ is an extrapolated DC amplitude, set by the LO power and TIA gain, and $v_{j,0}$ is the DC offset with no optical power incident (see Supplementary Materials for further details). The average phase was calculated as 
\begin{equation}
    \varphi_{\mathrm{avg}}=1/M\sum_{j=1}^{M}((1/A_{j})\braket{\hat{q}_{sq,j,k}}_{k}-v_{0}),
    \label{eq:AveragePhaseCalc}
\end{equation}
%$\varphi_{\mathrm{avg}}=1/M\sum_{j=1}^{M}((1/A_{j})\braket{\hat{q}_{sq,j,k}}_{k}-v_{0})$ 
where $\braket{\hat{q}_{sq,j,k}}_{k}$ refers to the mean of the 32,000 measured squeezed quadratures on spatial mode $j$. The average phase variance using entangled probe states was calculated as 
\begin{equation}
    \Delta^{2}\varphi_{\mathrm{avg}}=\langle\Delta(1/M\sum_{j=1}^{M}((1/A_{j})\hat{q}_{sq,j,k}-v_{j,0}))^{2}\rangle_{k},
    \label{eq:EntangledPhaseVar}
\end{equation}
%$\Delta^{2}\varphi_{\mathrm{avg}}=\braket{\Delta(1/M\sum_{j=1}^{M}((1/A_{j})\hat{q}_{sq,j,k}-v_{j,0}))^{2}}_{k}$, 
corresponding to the variance of the element-wise average of the quadratures across the spatial modes such that they are aligned in time. An example of the element-wise squeezed quadrature average is shown by the entangled trace in Fig. \ref{fig:3}(E), where the mean average DC offset and average phase estimator is marked by the dashed black line. Behind this trace is the element-wise shot noise quadrature average, which has a standard deviation of 1 in shot noise (SN) units. Therefore the entangled trace returns the DC offset with a precision \SI{0.197}{\decibel} below that of the shot noise.    

The average phase variance using separable probe states was calculated using 
\begin{equation}
    \Delta^{2}\varphi_{\mathrm{avg}}=\langle\Delta(1/M\sum_{j=1}^{M}((1/A_{j})\hat{q}_{sq,j,k+j}-v_{j,0}))^{2}\rangle_{k}
    \label{eq:SeparablePhaseVar}
\end{equation}
where this expression is identical to Eq.~\ref{eq:EntangledPhaseVar}, except the $k$ index of $\hat{q}_{sq}$ is offset by $j$ such that the temporal modes are no longer aligned in time, and the entanglement is destroyed. 
\begin{figure*}[!htb]
    \centering
    \includegraphics[width = \linewidth]{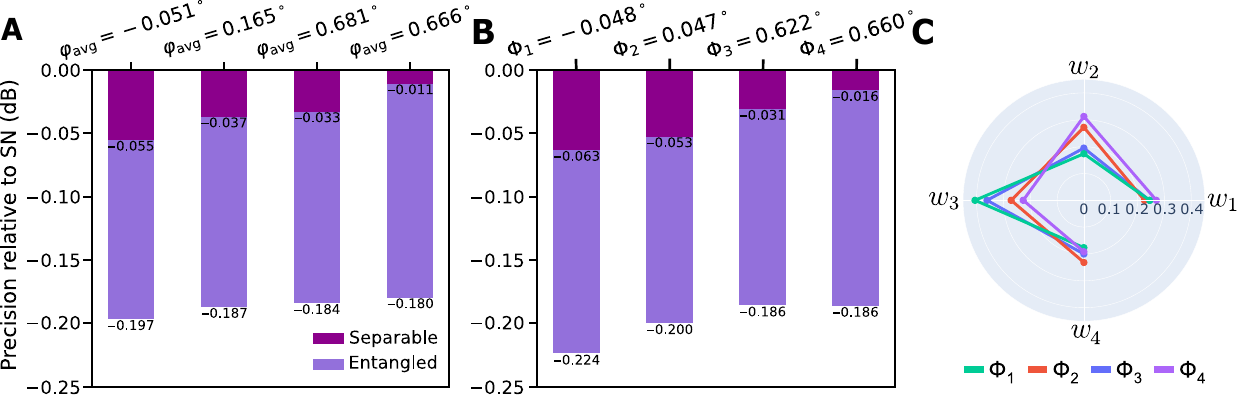}
    \caption{\textbf{Linear phase function sensing results} (A) Results of four different measurements of $\varphi_{\mathrm{avg}}$ with precisions relative to the shot noise (SN) labelled for separable and entangled sensing protocols. (B) Results from using the same data in (A) to sense the arbitrarily weighted linear function, $\Phi=\sum_{j=1}^{M}w_{j}\varphi_{j}$, with the values for $w_{j}$ given in (C).}
    \label{fig:4}
\end{figure*}

An example trace from which this was calculated is also shown in Fig.~\ref{fig:3}(E), with a standard deviation and average phase precision of \SI{0.029}{\decibel} below the shot noise. For the example given in Fig.~\ref{fig:3}(E), we find that $\varphi_{\textrm{avg}}=0.165^{\circ}$. The measurement was repeated another three times, for a variety of DC offsets at each detector with the resulting $\varphi_{\textrm{avg}}$ and separable and entangled precisions shown in Fig.~\ref{fig:4}(A). The entangled sensing protocol measured $\varphi_{\textrm{avg}}$ with precisions between \SI{0.180}{\decibel} and \SI{0.197}{\decibel} below the SN, whereas the separable state protocol obtained precisions between \SI{0.011}{\decibel} and \SI{0.055}{\decibel} below the SN. These results show that by maintaining the effective noise cancelling, cross-spatial mode entanglement achieves a precision that consistently outperforms that of the separable states, where the precision relative to that of the shot noise is simply limited to the squeezing that exists at each mode.

As well as $\varphi_{\textrm{avg}}$, we can also consider constructing an estimator from an arbitrarily weighted sum of the measured modes $\Phi=\sum_{j=1}^{M}w_{j}\varphi_{j}$ where $\sum_{j=1}^{M}w_{j}=1$. This is particularly relevant since the optimal weights for each term in the sum should match the splitting ratio of the squeezing across the spatial modes assuming the loss on each mode is the same \cite{Zhuang2018, Gatto2019}. Due to imperfections in the BSN, varying detection efficiencies and noise profiles between the detectors, we see from Fig.~\ref{fig:3}(D) that the squeezing is not equally split across the four modes indicating that a higher performance can be found by interrogating an estimator beyond the simple mean. For each of the four sensed phases shown in Fig.~\ref{fig:4}(A) we ran a differential evolution optimiser to determine the values of $w_{j}$ that gave rise to the optimal measurement precision and obtained the results shown in Fig.~\ref{fig:4}(B). The entangled sensing protocol measured the arbitrarily weighted linear phase function with precisions between 0.186 and 0.224 dB below the SN, whereas the separable state protocol obtained precisions between 0.016 and 0.063 dB below the SN. In all cases we find that increased precision is obtained for the weighted sum compared to the simple mean. The values of $w_{j}$ for each of the four cases is shown in Fig.~\ref{fig:4}(C). 

\section{Discussion}
\label{Discussion}

In this work we have demonstrated entanglement-enhanced measurements of various linear functions of TOPM phase shift parameters in an SOI PIC. The continuous variable entanglement was generated on-chip from a squeezed vacuum state incident on an integrated four-mode BSN, with the state tomography and phase sensing implemented with an array of four integrated homodyne detectors. We have shown we are able to measure a weighted linear function with a precision of 0.199(16)~dB below the SN limit, whereas the separable state sensing equivalent achieves a precision of 0.041(18)~dB below the SN.

Where all previous DQS experiments have been implemented with bulk optics, fundamentally limiting the number of modes that can be practically achieved, integrated photonics provides an inherently scalable and repeatable alternative platform. The most significant current limitation of the platform, as illustrated by this experiment, is the losses introduced by coupling the squeezed light on or off the chip where the greatest contribution to the total system efficiency is the ~2.5 dB grating coupler loss. Constructing a fully integrated squeezing source, BSN and homodyne detector array could remove the coupling losses, as well as the need for active phase locking \cite{green2026generation, chen2026heterogeneouslyintegratedsqueezedlightgeneration}.   

The TOPMs used to implement the phase parameters in this experiment are representative of real world samples, but future iterations of this PIC could feature regions of suspended waveguide to load samples for entanglement-enhanced sensing of average sample or material properties, for example biological, chemical or medical substances. This then leads us to stress the advantage of this fully room temperature demonstration, where many biological samples would not be able to survive cryogenic temperatures, and further highlights a route towards robust and portable entanglement-enhanced sensors.

\subsection{Acknowledgements}
The authors thank Mikkel Larsen, Virginia D'Auria, Edward Deacon, Matthew Stafford and Rowan Hoggarth for valuable scientific discussions and insights.\\\\
This work was supported by the European Research Council starting grant ``PEQEM'' (ERC-2018-STG 803665), the EPSRC Fellowship (EP/M024385/1), the EPSRC grant ``Mono-Squeeze'' (EP/X016218/1), the Quantum Position, Navigation and Timing (QEPNT) Hub (EP/Z533178/1), the Quantum Sensing, Imaging and Timing (QuSIT) Hub (EP/Z533166/1), and the Integrated Quantum Networks (IQN) Hub (EP/Z533208/1). O.M.G. \& B.P. acknowledge support from EPSRC Quantum Engineering Centre for Doctoral Training (EP/S023607/1). A.S.C. acknowledges support from The Royal Society (URF/R/221019, RF/ERE/210098, RF/ERE/221060). J.C.F.M is grateful for support from his Philip Leverhulme Prize.
\subsection{Author contributions}
B.P, G.F and J.C.F.M formulated the scheme for DQS for the PIC designed by J.F.T. B.P constructed the setup, assembled the PCB, wirebonded the PIC and tested the setup with input from O.M.G, R.N.C and B.D.J.S. The squeezed light source was assembled by J.F and tested by B.P and O.M.G. B.P, T.E and O.M.G. designed and constructed the noise locking setup. The entanglement verification and phase sensing data was collected and analysed by B.P. G.F, A.S.C and J.C.F.M provided supervision for the project. B.P, R.N.C, O.M.G and A.S.C prepared the manuscript with input from all authors.

\bibliography{reference}% Produces the bibliography via BibTeX.

\end{document}